\documentclass[12pt]{cta-author}

\usepackage[T1]{fontenc}
\usepackage[english]{babel}
\usepackage{array}
\usepackage{calc}
\usepackage{amsmath}
\usepackage{amsthm}
\usepackage{amssymb}
\usepackage{bm}
\usepackage{graphicx}
\usepackage{natbib}
\usepackage{epstopdf}
\usepackage{algorithm}
\usepackage{subfigure}
\usepackage{color}

\usepackage{enumitem}
\usepackage{multirow}
\usepackage{hhline}
\usepackage{float}
\usepackage{makecell}
\usepackage{boldline}

{}
{}
{}

\providecommand{\tabularnewline}{\\}

\begin{document}
	
	\supertitle{Submission Template for IET Research Journal Papers}
	
	\title{Detecting Load Redistribution Attacks via Support Vector Models}
	
	\author{\au{Zhigang Chu$^{1\corr}$}, \au{Oliver Kosut$^{1}$}, \au{Lalitha Sankar$^{1}$}}
	
	\address{\add{1}{School of Electrical, Computer and Energy Engineering, Arizona State University, Tempe, AZ, 85287, USA}
		\email{zchu2@asu.edu}}
	
	\begin{abstract}
		A machine learning-based detection framework is proposed to detect a class of cyber-attacks that redistribute loads by modifying measurements. The detection framework consists of a multi-output support vector regression (SVR) load predictor that predicts loads by exploiting both spatial and temporal correlations, and a subsequent support vector machine (SVM) attack detector to determine the existence of load redistribution (LR) attacks utilizing loads predicted by the SVR predictor. Historical load data for training the SVR are obtained from the publicly available PJM zonal loads and are mapped to the IEEE 30-bus system. The SVM is trained using normal data and randomly created LR attacks, and is tested against both random and intelligently designed LR attacks. The results show that the proposed detection framework can effectively detect LR attacks. Moreover, attack mitigation can be achieved by using the SVR predicted loads to re-dispatch generations.
	\end{abstract}
	
	\maketitle
	
	\section{Introduction\label{sec:Introduction}}
	%
	%
	%
	%
	Leveraging information technology, the operation of modern electric power grids largely rely on real-time sensing, monitoring, communication, and control. State estimation (SE) utilizes the power system measurements collected by the supervisory control and data acquisition (SCADA) system to estimate the operating states. These states are used by the energy management system (EMS) to allow for real-time control of power system. In the last decade, the cyber-security of SE has been studied with considerable attention. A class of false data injection (FDI) attacks that replace measurements with counterfeits have been shown to be able to easily spoof SE and the traditional bad data detector (BDD) \cite{Liu2009}. This finding serves as the basis of a wide class of FDI attacks, called load redistribution (LR) attacks, which make it appear as if the loads are redistributed among load buses while the total load remain unchanged.
	
	Worst-case consequences of LR attacks can be found using bi-level optimization problems. These attacks can be designed to cause physical or economic consequences. For physical consequences, \cite{Zhang2016TSG} attempts to find an attack to mask the outage of a transmission line, and \cite{Liang2015} designs attacks that can cause physical overflows. For economic consequences, \cite{Moslemi2018} and \cite{Jia2014} show that LR attacks can change locational marginal prices, and/or make profit for attackers. Therefore, it is crucial to develop techniques to detect and mitigate LR attacks.
	
	Various attack detection techniques have been presented in the literature. In \cite{An2019}, the authors propose a multivariate Gaussian-based anomaly detector trained using correlation features of micro phasor measurement units ($\mu$PMUs), but this detector requires installation of $\mu$PMUs in the system. Liu \textit{et al.} \cite{Liu2018} detect and identify attacks using reactance perturbation, but this method only works when the attacker has limited resources. The authors of \cite{Che2019} attempt to mitigate LR attacks using a tri-level optimization approach, and the authors of \cite{Li2019} try to identify LR attacks by monitoring abnormal load deviations and suspicious branch flow changes. However, they both only focus on attacks that cause line overflows. In \cite{Liu2019}, a financially motivated FDI attack model is analyzed and a robust incentive-reduction strategy is proposed to deter such attacks by protecting a subset of meters.  More generally, machine learning techniques are also deployed in detecting LR attacks. For example, \cite{Ozay2016} proposes supervised and semi-supervised machine learning algorithms to detect FDI attacks by exploiting the relationships between statistical and geometric properties of attack vectors employed in the attack scenarios. A deep reinforcement learning-based approach is proposed to detect LR attacks in \cite{An2019a}. In \cite{Pinceti2018}, three machine learning techniques are introduced for attack detection, namely nearest neighbor, semi-supervised one class SVM, and replicator neural network. These three algorithms compare estimated loads with historical loads and use thresholding to determine the existence of LR attacks.
	
	\emph{Estimation-Detection Framework:}
	In this paper, we introduce an LR attack detection framework based on support vector models by leveraging the historical load information commonly available to system operators. Unlike most existing approaches in the literature, our method determines the existence of LR attacks directly through the estimated loads, without requiring installations of new devices nor protection of specific measurements. When an LR attack occurs, the estimated loads obtained from the SE results are different from the true loads, but the net loads are the same. Thus, if accurate load predictions are available, the existence of LR attacks can be determined by comparing the predicted and estimated loads. Moreover, if an LR attack is detected, the predicted loads can be directly used to re-dispatch generation instead of using the estimated loads. By doing this, the attack consequences can be temporarily mitigated, giving operators time to perform other corrective actions.
	
	\emph{Support Vector Models:}
	In particular, we propose a support vector regression (SVR) \cite{Smola2004} based load predictor to accurately predict loads, and a subsequent support vector machine (SVM) \cite{Cortes1995} based attack detector that compares the predicted and observed loads to detect LR attacks. Our choice of this modular design aims to separate the prediction and classification, so that each module can be independently enhanced (\textit{e.g.,} using additional features) and also replaced by other methods, as seen fit. Support vector models are optimization-based machine learning approaches that can be used for both regression and classification purposes. There are many different machine learning methods, and we choose support vector models for the following reasons: (i) they are mature methods that have been proven to be effective for various regression/classification tasks in power systems, including transient stability assessment \cite{Yuanhang2015}, component outage estimation \cite{Eskandarpour2017}, and state estimation \cite{Kirincic2019}; (ii) they are analytically developed models with fewer and easier to tune parameters compared to many other machine learning methods, \textit{e.g.,} neural networks. 
	
	SVR has been widely used for load prediction in electric power systems. In \cite{Qiang2019}, a short-term load forecasting algorithm is proposed combining SVR and particle swarm optimization. The authors of \cite{Capuno2017} proposes a SVR model that predicts very short term loads using weather data and day ahead predicted loads as features. Similar features along with additional time-related features are used to train a SVR model that predicts short term and mid term loads in \cite{Su2017}. In \cite{Azad2018}, Azad \textit{et al,} predict the daily peak load using the historical peak load consumption and the corresponding temperature and relative humidity. Chong \textit{et al,} propose a K-step ahead prediction using SVR in \cite{Chong2017}.
	
	\textit{Proposed SVR Load Predictor:} The aforementioned references focus on predicting the net load utilizing temporal correlation. To the best of our knowledge, we are one of the first to predict loads at each bus using SVR, leveraging both spatial and temporal correlations between all the loads in the system. Features selected for the SVR predictor include historical load values of all loads chosen at distinct time intervals 
	prior to the target time (\textit{e.g.}, one hour before, one day before, etc.) as well as the specific time information  (\textit{e.g.,} month, weekday/weekend).
	This choice allows for conveniently using the same features to predict loads at different buses as the temporal features for all loads implicitly capture the spatial correlations among them. 
	
	\textit{Proposed SVM Detector:} SVM is a supervised learning approach to solve classification problems,  based on learning separating hyperplanes. Our approach using SVR to detect attacks largely mirrors existing approaches; our key contribution is in how we generate the training data needed to learn the SVM model to classify accurately over a large class of attacks. We now describe the dataset and our approach to train and test the two models.
	
	\emph{Dataset:} We train and test our models using the publicly available PJM metered zonal load data \cite{PJM2019}. We map each of the 20 zones of the PJM data to a load bus in the IEEE 30-bus system, leveraging the fact that there are 20 loads in this system.
	
	\emph{Training and Testing:}
	To apply SVM on attack detection, it is necessary to create training data in both classes, namely \textit{normal} and \textit{attacked} data. The SVR predicted loads and the true loads (assuming trustworthy historical data) naturally form the normal data. 
	For the attacked data, we propose a novel approach that generates random LR attacks in order to maximally explore the attack space, and thereby enhance accuracy in detecting \textit{any} LR attack. Each of these attacks alters a random number of loads, and a Gaussian distribution is used to generate the deviation of each load from its true value. The severity of the attacks is controlled by varying the maximum deviation percentage over all loads. Our approach also guarantees the net load change is 0 to satisfy the constraints of LR attacks. We use 80\% of the data for training, and the remaining 20\% for testing. 
	
	In addition to the random attacks, we also generate two types of intelligently designed LR attacks, namely cost maximization (CM) and line overflow (LO) attacks, to test the effectiveness of our SVM attack detector. CM attacks aim to maximize the operation cost \cite{Yuan11}; and LO attacks attempt to overflow a target transmission line \cite{Liang2015}. These two types of attacks are designed through optimizations to maximize their economic/physical consequences. 
	
	Our results show that the proposed attack estimation-detection framework can effectively predict and detect both random and intelligently designed LR attacks. Moreover, we illustrate that using the SVR predicted loads to re-dispatch when attacks are detected can significantly reduce the attack consequences.
	
	
	\textit{Summary of Contributions:} 
	The key contributions of this paper are as follows:
	
	1. We propose an LR attack detection framework consisting of an SVR load predictor and a subsequent SVM attack detector. This modular design enables separate enhancement of each block, and also provides sufficiently accurate predicted loads for attack mitigation purposes. 
	
	2. The SVR predictor leverages both temporal and spatial correlations within the historical load data to allow for prediction of bus-level loads. Through training and testing the proposed SVR predictor on the PJM metered load data \cite{PJM2019}, we show that it can accurately predict every load in the system. 
	
	3. Utilizing the SVR predicted loads, we train the SVM detector using normal data and random LR attacks designed to maximally explore the attack space. 
	
	4. The performance of the detection framework is tested on random attacks as well as two types of intelligently designed LR attacks. These attacks aim to cause economic/physical consequences. Our simulation results show that our detection framework can significantly reduce the impact of LR attacks.
	
	
	The rest of this paper is organized as follows. Section \ref{sec:LR_Attacks} introduces LR attacks and existing approaches to create intelligently designed LR attacks. Section \ref{sec:ProposedDetector} describes the structure of the proposed attack detection framework, the formulations of SVR and SVM, as well as random LR attack creation method for SVM training purpose. Section \ref{sec:Results} illustrates the performance of the SVR load predictor and the SVM attack detector. Concluding remarks are presented in Section \ref{sec:conclusion}.
	
	\section{Load Redistribution Attacks \label{sec:LR_Attacks}}
	\subsection{Load Redistribution (LR) Attacks and Unobservable False Data Injection (FDI) Attacks}
	\textit{Definition 1:} LR attacks are a class of cyber-attacks that redistribute loads among the buses, while keeping the net load unchanged. The false loads in an LR attack $\boldsymbol{P}_{\text{Atk}}$ satisfies
	\begin{flalign}
	& \boldsymbol{P}_{\text{Atk}}=\boldsymbol{P}+\Delta\boldsymbol{P},  \label{eq:LRAtk1}\\
	& \sum_{i} \Delta{P_{i}}=0,\label{eq:LRAtk2}
	\end{flalign}
	where $\boldsymbol{P}$ is the true load vector, $\Delta\boldsymbol{P}$ is the load change caused by attack, and $i$ is the load index.
	
	\textit{Definition 2:} The load shift $\tau$ is defined to be the largest load change in percentage of the true loads:
	\begin{equation}
	\tau = \underset{i}{\max} \left\vert\frac{\Delta{P_{i}}}{P_{i}}\right\vert\times 100\%. \label{eq:loadshift}
	\end{equation}
	We use $\tau$ as an intrinsic metric to characterize the detectability of LR attacks. We found that it is trivial to detect attacks with sufficiently large $\tau$, because load deviations far from true values are suspicious. Thus, an attacker is likely to limit $\tau$ to avoid detection by this metric. In this paper, we only consider LR attacks with $\tau \le 20\%$.
	
	The most common way to generate LR attacks in the literature is through unobservable FDI attacks against power system state estimation (SE). FDI attacks are a class of cyber-attacks that involves an attacker maliciously replacing power system measurements with counterfeits. Under DC power flow assumption\footnote[1]{For simplicity, we focus on DC power flow settings, but our work can be generalized to AC cases as in \cite{Liang2015}.}, the true measurement vector $\textbf{z}$, consisting of the line power flow and bus power injection measurements, is given by 
	\begin{equation}
	\textbf{z}=\boldsymbol{H\theta}+\boldsymbol{e}, \label{eq:trueMeas}
	\end{equation} 
	where $\boldsymbol{\theta}$ is the state vector (voltage angles), $\boldsymbol{H}$ is the dependency matrix between measurements and states, and $\boldsymbol{e}$ is the noise vector. 
	
	\textit{Definition 3:} A false measurement vector $\bar{\textbf{z}}$ created with state attack vector $\textbf{c}$, 
	\begin{equation}
	\bar{\textbf{z}} = \boldsymbol{H}(\boldsymbol{\theta}+\textbf{c})+\boldsymbol{e}, \label{eq:unobservableMeas}
	\end{equation}
	is \textit{unobservable} to the conventional bad data detector (BDD) embedded with SE, because it is not distinguishable from the true measurements if the true states were $(\boldsymbol{\theta}+\textbf{c})$. 
	
	Let $\boldsymbol{B}$ be the dependency matrix between bus power injections and states, and let $\boldsymbol{G}$ be a given generation vector, then the bus power injections without attack can be expressed as 
	\begin{equation}
	\boldsymbol{G}-\boldsymbol{P}=\boldsymbol{B\theta}. \label{eq:injectionWithoutAtk}
	\end{equation}
	With attack, the false injections are given by
	\begin{equation}
	\boldsymbol{G}-\boldsymbol{P}_{\text{Atk}}=\boldsymbol{B}(\boldsymbol{\theta}+\boldsymbol{c}). \label{eq:injectionWithAtk}
	\end{equation}
	Substituting \eqref{eq:injectionWithoutAtk} into \eqref{eq:injectionWithAtk} yields the load change vector 
	\begin{equation}
	\Delta\boldsymbol{P}=\boldsymbol{P}_{\text{Atk}} - \boldsymbol{P} = - \boldsymbol{Bc}.\label{eq:falseLoads}
	\end{equation}
	Note that since $\boldsymbol{1}^T\boldsymbol{B}=\boldsymbol{0}^T$, the net load change is $\sum\limits_{i} \Delta{P_{i}}=-\boldsymbol{1}^T\boldsymbol{Bc}=0$. Thus, given a generation dispatch, an unobservable FDI attack leads to an LR attack.
	
	\subsection{Intelligently Designed LR Attacks\label{sec:designedAtk}}
	Although an attacker can inject arbitrary $\boldsymbol{c}$ as long as it controls the measurements corresponding to all non-zero entries of $\boldsymbol{Hc}$, its goal will be to maliciously choose $\boldsymbol{c}$ so that the resulting false loads can mislead the system re-dispatch to cause physical and/or economical consequences. We define these attacks as \textit{intelligent attacks}, whose consequences can be maximized by solving optimization problems. In this paper, we consider two specific intelligent attacks to test the robustness of our proposed detector, namely cost maximization (CM) attacks \cite{Yuan11} and line overflow (LO) attacks \cite{Liang2015}.
	
	CM attacks are a class of FDI attacks that aim to maximize the operation cost after re-dispatch. The attack vector $\boldsymbol{c}$ of CM attacks can be obtained through the following bi-level optimization problem:
	\begin{subequations}\label{eq:CMAtk}
		\begin{flalign}
		\underset{\boldsymbol{c}}{\text{maximize}}\: \hspace{0.2cm} & \boldsymbol{a}^T \boldsymbol{G}^*\label{eq:CMObj}\\
		\text{subject to}\hspace{0.2cm}\; & -\tau \boldsymbol{P}\le \boldsymbol{Bc}\le \tau \boldsymbol{P}\label{eq:CM_Con_loadshift}\\
		& \left\{\boldsymbol{G}^{*}, \boldsymbol{P_L}^*\right\} =\text{arg}\left\{ \underset{\boldsymbol{G, P_L}}{\text{min}}\: \boldsymbol{a}^T\boldsymbol{G}\right\} \label{eq:OPFObj_MINCOST}\\
		& \text{subject to}\hspace{0.2cm}\; \sum \boldsymbol{G} = \sum \boldsymbol{P}\label{eq:OPFCon_PowerBalance}\\
		& \hspace{1.7cm} \boldsymbol{P_L} = \boldsymbol{R}(\boldsymbol{G-P+Bc}) \label{eq:OPFCon_PL}\\
		& \hspace{1.7cm} -\boldsymbol{P_L}^{\max} \le \boldsymbol{P_L} \le \boldsymbol{P_L}^{\max} \label{eq:OPFCon:PL_limit}\\
		& \hspace{1.7cm} \boldsymbol{G}^{\min} \le \boldsymbol{G} \le \boldsymbol{G}^{\max} \label{eq:OPFCon:Gen_limit}
		\end{flalign}
	\end{subequations}
	where $\boldsymbol{a}$ is the generation cost, $\boldsymbol{P_L}$ is the cyber line power flows, $\boldsymbol{R}$ is the power transfer distribution factor (PTDF) matrix, $\boldsymbol{P_L}^{\max}$ is the line power flow limits, and $\boldsymbol{G}^{\max}$ and $\boldsymbol{G}^{\min}$ are generation upper and lower limits, respectively. In the upper level, \eqref{eq:CMObj} models the attacker's objective to maximize the operation cost, and \eqref{eq:CM_Con_loadshift} models the load shift limit. The lower level problem \eqref{eq:OPFObj_MINCOST}-\eqref{eq:OPFCon:Gen_limit} is the system DCOPF under attack. This bi-level optimization problem can be converted to a single level mixed-integer linear program (MILP) by replacing the lower level DCOPF with its Karush-Kuhn-Tucker (KKT) conditions \cite{BoydBook}, and then converting the complementary slackness conditions to mixed integer constraints. The optimal $\boldsymbol{c}$ is obtained by solving the MILP. 
	
	LO attacks attempt to maximize the physical power flow on a target line $l$ after re-dispatch, and possibly cause overflows. Optimal $\boldsymbol{c}$ for LO attacks can be obtained by changing the objective function of \eqref{eq:CMAtk} to maximizing physical power flow:
	\begin{flalign}
	\underset{\boldsymbol{c}}{\text{maximize}}\: \hspace{0.2cm} &\left\vert\boldsymbol{P_{L}}^{l*}-\boldsymbol{R}_l\cdot \boldsymbol{Bc}\right\vert \label{eq:LOObj}\\
	\notag \text{subject to}\hspace{0.2cm}\; & \eqref{eq:CM_Con_loadshift} - \eqref{eq:OPFCon:Gen_limit},
	\end{flalign}
	where $\boldsymbol{P_{L}}^{l*}$ is the optimal cyber power flow on target line $l$, $\boldsymbol{R}_l$ is the $l^{\text{th}}$ row of $\boldsymbol{R}$, and the second term in \eqref{eq:LOObj} characterizes the impact of false loads on the physical power flow of line $l$.
	
	\section{Proposed Attack Detection Framework \label{sec:ProposedDetector}}
\begin{figure}[h]
		\centering{}\includegraphics[trim=0 0.2cm 0 0.6cm, scale=0.6]{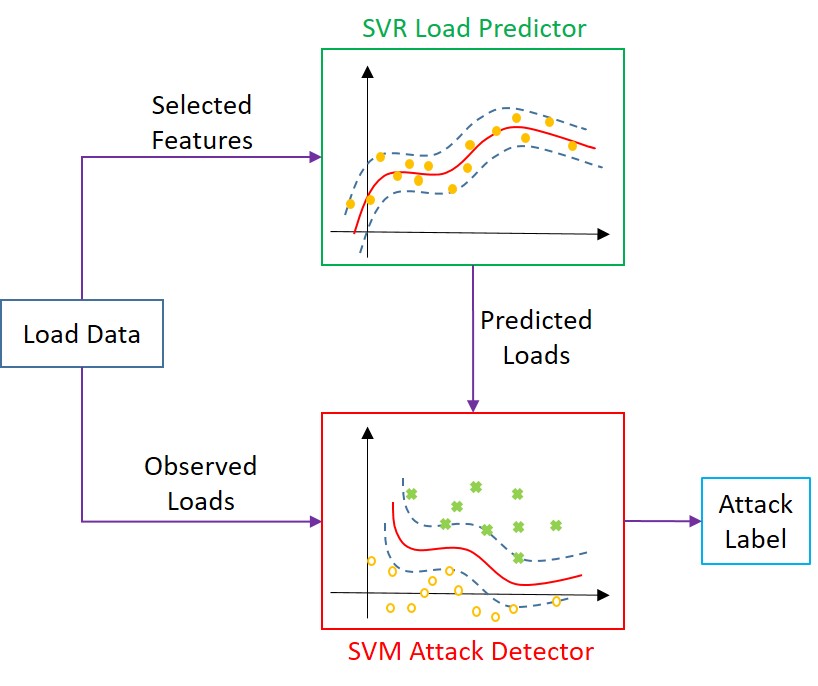}\protect\protect\caption{Structure of the proposed LR attack detection framework. \label{fig:structure}}
	\end{figure}

	Figure \ref{fig:structure} illustrates the structure of our proposed LR attack detection framework. During the real-time operation, features are selected from the historical load data until the current time step to capture both spatial and temporal correlations. Loads at the next time step are then predicted by the SVR load predictor using these features. One SVR model is trained for each load using the same features. Subsequently, the SVM attack detector takes the predicted loads and loads estimated after SE to determine the existence of LR attacks. 
	
	For detecting attacks, it should suffice to skip the SVR load predictor and plug all SVR features into the SVM to perform classification. However, in this paper we include the SVR for the following two reasons. The first one is that we aim to not only find an attack detection technique, but also have a corrective mechanism when attacks are detected. Using the (accurate) predicted loads to perform control actions when attacks are flagged provides time to locate the attacked measurements without causing severe consequences. The second reason is for easier scaling of the proposed models to large-scale power systems. Without the SVR predictor, the number of features used in SVM classifier will be very large, making it difficult to train and perform real-time classifications. With the SVR predictor in place, the SVM detector only needs the predicted and observed load values as features, making it useful for large-scale systems.

	\subsection{The SVR Load Predictor\label{sec:trainSVR}}
	Given data samples $\boldsymbol{x}_j\in \mathbb{R}^p, j=1,2,3,...,m$ and target values $\boldsymbol{y}\in \mathbb{R}^m$, an SVR attempts to find the best parameters $\boldsymbol{w}_r$ and $b_r$ to fit $|y_j-\boldsymbol{w}_r^T\phi(\boldsymbol{x}_j)-b_r|\le \varepsilon$ by solving the following optimization problem \cite{Smola2004}: 
	\begin{subequations} \label{eq:SVR}
		\begin{flalign}
		\hspace{-0.3cm}\underset{\boldsymbol{w}_r,b_r,\zeta_j,\zeta_j^{\prime}}{\text{minimize}}\: \hspace{0.2cm} & \frac{1}{2}\boldsymbol{w}_r^T\boldsymbol{w}_r + M\sum\limits_{j=1}^{n}(\zeta_j+\zeta_j^{\prime})\label{eq:Obj_SVR}\\
		\text{subject to}\hspace{0.2cm}\; & y_j - \boldsymbol{w}_r^T\phi(\boldsymbol{x}_j)-b_r\le \varepsilon + \zeta_j \hspace{0.3cm}(\alpha_j)\label{eq:posConSVR}\\
		& \boldsymbol{w}_r^T\phi(\boldsymbol{x}_j)+b_r-y_j \le \varepsilon + \zeta_j^{\prime} \hspace{0.3cm}(\alpha_j^{\prime})\label{eq:negConSVR}\\
		& \zeta_j, \zeta_j^{\prime} \ge 0, \forall j, \label{eq:zetaConSVR}
		\end{flalign}
	\end{subequations}
	where $\varepsilon$ is the regression tolerance, $\zeta_j, \zeta_j^{\prime}$ are slack variables to allow for outliers, $M$ is the penalty factor for outliers, $\alpha_j, \alpha_j^{\prime}$ are dual variables, and $\phi(\cdot)$ is a function that implicitly maps the data samples to a higher dimensional space. The dual formulation has a smaller number of variables and allows for applying the kernel trick:
	\begin{subequations}\label{eq:SVR_dual}
		\begin{flalign}
		\underset{\boldsymbol{\alpha,\alpha^{\prime}}}{\textnormal{minimize}}\hspace{0.3cm}&\frac{1}{2}(\boldsymbol{\alpha-\alpha^{\prime}})^T\boldsymbol{Q}(\boldsymbol{\alpha-\alpha^{\prime}})\\
		\notag & +\varepsilon \boldsymbol{1}^T(\boldsymbol{\alpha+\alpha^{\prime}})-y^T(\boldsymbol{\alpha-\alpha^{\prime}})\\
		\textnormal{subject to}\hspace{0.2cm}& \boldsymbol{1}^T(\boldsymbol{\alpha-\alpha^{\prime}})=0\\
		&0\le \alpha_j,\alpha_j^{\prime}\le M, \forall j
		\end{flalign}
	\end{subequations}
	where $\boldsymbol{Q}$ is a positive semi-definite matrix, and $Q_{ij}=Q(\boldsymbol{x}_i,\boldsymbol{x}_j)=\phi(\boldsymbol{x}_i)^T\phi(\boldsymbol{x}_j)$ is the kernel. Once the optimal solutions ($\boldsymbol{\alpha}^*, \boldsymbol{\alpha}^{\prime*}$) are obtained, the regression value $y_{\text{new}}$ of a new data sample $\boldsymbol{x}_{\text{new}}$ can be computed as
	\begin{equation}
	y_{\text{new}} = \sum_{j=1}^{n}(\alpha_j^*-\alpha_j^{\prime*})Q(\boldsymbol{x}_j, \boldsymbol{x}_{\text{new}}). \label{eq:SVR_recover}
	\end{equation}
	
	To accurately predict the load values, many different features can be used, including time, weather, temperature, location, and load type (residential/commercial/industrial). Intuitively, it would be the best if we use all the features to perform the prediction, but many of them are unavailable, and some of them may be redundant. The features used in the SVR load predictor also depend on the available dataset. For example, the time step of the prediction depends on how frequently the historical load data are recorded. For the specific dataset we use in this paper, we select time information and historical load values at several time points relative to the target time to capture the temporal correlation, and load values at the same time points for all loads to capture the spatial correlation. Details of selected features for the SVR load predictor will be given in Section \ref{sec:SVR_Perform}.
	
	\subsection{The SVM Attack Detector\label{sec:trainSVM}}
	Given data samples $\boldsymbol{u}_j\in \mathbb{R}^q, j=1,2,3,...n$ and a vector of class labels $\boldsymbol{v}\in \{1,-1\}^n$, an SVM attempts to find the decision boundary with the maximal margin to best classify $\boldsymbol{u}_j$ by solving the following optimization problem \cite{Cortes1995}:
	\begin{subequations} \label{eq:SVM}
		\begin{flalign}
		\hspace{-0.3cm}\underset{\boldsymbol{w}_m,b_m,\lambda_j}{\text{minimize}}\: \hspace{0.2cm} & \frac{1}{2}\boldsymbol{w}_m^T\boldsymbol{w}_m + C\sum\limits_{j=1}^{n}\lambda_j\label{eq:Obj_SVM}\\
		\text{subject to}\hspace{0.2cm}\; & {v}_j(\boldsymbol{w}_m^T\phi(\boldsymbol{u}_j)+b_m)\ge 1 - \lambda_j \hspace{0.3cm} (\beta_j)\label{eq:ConSVM}\\
		& \lambda_j \ge 0, \forall j. \label{eq:zetaConSVM}
		\end{flalign}
	\end{subequations}
	Similar to the SVR formulation in \eqref{eq:SVR}, $\lambda_j$ is a slack variable to allow for outliers, $C$ is its penalty factor, and $\beta_j$ is the dual variable. Again, applying the kernel trick, the dual formulation is used:
	\begin{subequations}\label{eq:SVM_dual}
		\begin{flalign}
		\label{eq:SVM_Dual}\underset{\boldsymbol{\beta}}{\text{minimize}} \hspace{0.3cm} & \frac{1}{2}\boldsymbol{\beta}^T\boldsymbol{Q\beta}-\boldsymbol{1}^T\boldsymbol{\beta}\\
		\text{subject to} \hspace{0.3cm} & \boldsymbol{v}^T\boldsymbol{\beta} =0\\
		& 0\le \beta_j \le C, \forall j.
		\end{flalign}
	\end{subequations}
	Note that here $Q_{ij}=v_iv_jQ(\boldsymbol{u}_i, \boldsymbol{u}_j)=v_iv_j\phi(\boldsymbol{u}_i)^T\phi(\boldsymbol{u}_j)$. Once the optimal solution $\boldsymbol{\beta}$ is acquired, the label $v_{\text{new}}$ for a new input data sample $\boldsymbol{u}_{\text{new}}$ can be obtained by 
	\begin{equation}
	v_{\text{new}} = \text{sgn}(\sum_{j=1}^{n}v_j\beta_j^*Q(\boldsymbol{u}_j, \boldsymbol{u}_{\text{new}})) \label{eq:SVM_recover}
	\end{equation}
	where $\text{sgn}(\cdot)$ is the sign function. The features in $\boldsymbol{u}_j$ include the SVR predicted loads, the observed loads, and the same time information used in the SVR.
	
	\subsection{Generating Random LR Attacks to Train the SVM\label{sec:randomAtk}}
	We train the SVM detector using normal data and randomly designed LR attacks. The SVM detector trained using random attacks is expected to maximally explore the space of LR attacks, and hence, perform well in detecting any LR attacks. Given true loads $\boldsymbol{P}$, the false loads $\boldsymbol{P}_{\text{Atk}}$ in these random attacks are acquired using \eqref{eq:LRAtk1}, $\boldsymbol{P}_{\text{Atk}}=\boldsymbol{P}+\Delta \boldsymbol{P}$. Thus, finding $\boldsymbol{P}_{\text{Atk}}$ is equivalent to finding $\Delta \boldsymbol{P}$. In each attack, we assume the attacker changes $K$ loads at random, whose indices form a set $\mathcal{K}$, so that $\Delta P_{\mathcal{K}(k)}$ indicates the load change of the $k^{\text{th}}$ attacked load, $k=1,2,\dots,K$. The load changes of these attacked loads, denoted $\boldsymbol{\gamma}$, can be arbitrary. However, according to the LR attack property \eqref{eq:LRAtk2}, they must be constrained to have a 0 sum. Thus, we model $\boldsymbol{\gamma}$ with a joint Gaussian distribution with 0 mean and covariance matrix $\boldsymbol{\Gamma}$:
	\begin{flalign}
	\boldsymbol{\gamma} & \sim \mathcal{N}(\boldsymbol{0, \Gamma}) \label{eq:randomGamma} \\ 
	\gamma_k & = \Delta P_{\mathcal{K}(k)}. \label{eq:gammak}
	\end{flalign}
	Given a load shift $\tau$, the diagonal entries of $\boldsymbol{\Gamma}$ must satisfy
	\begin{equation}
	\Gamma_{kk} = Var(\gamma_k) = (\frac{1}{2}\tau P_{\mathcal{K}(k)})^2, \forall k \label{eq:Gammakk}
	\end{equation}
	to ensure that the probability of $|\gamma_k|\le \tau P_{\mathcal{K}(k)}$ is 95\%, because the probability of deviating beyond $2\times$standard deviation in a Gaussian distribution is 5\%. Recall that the load changes caused by a valid LR attack must satisfy \eqref{eq:LRAtk2}, which can be rewritten as 
	\begin{equation}
	\sum\limits_{i}\Delta P_i=\sum\limits_{k}\Delta P_{\mathcal{K}(k)}=\boldsymbol{1}^T\boldsymbol{\gamma}=0. \label{eq:sumDeltaP}
	\end{equation}
	Eq. \eqref{eq:sumDeltaP} is equivalent to
	\begin{flalign}
	\notag	E[(\boldsymbol{1}^T\boldsymbol{\gamma})^2]&=E[\boldsymbol{1}^T\boldsymbol{\gamma}\boldsymbol{\gamma}^T\boldsymbol{1}]\\
	\notag &=\boldsymbol{1}^T\boldsymbol{\Gamma} \boldsymbol{1}\\
	&=0. \label{eq:sumGamma}
	\end{flalign}
	Finding a valid $\boldsymbol{\gamma}$ is equivalent to finding a positive semidefinite matrix $\boldsymbol{\Gamma}$ that satisfies \eqref{eq:Gammakk} and \eqref{eq:sumGamma}. Since $\boldsymbol{\Gamma}$ is a covariance matrix, it must be positive semidefinite: 
	\begin{equation}
	\boldsymbol{\Gamma} \succeq 0. \label{eq:Sdetection probability}
	\end{equation}
	Any $\boldsymbol{\Gamma}$ satisfying \eqref{eq:Gammakk}, \eqref{eq:sumGamma} and \eqref{eq:Sdetection probability} would suffice for our application. Finding $\boldsymbol{\Gamma}$ is equivalent to solving a semidefinite program with arbitrary objective, constrained by \eqref{eq:Gammakk}, \eqref{eq:sumGamma} and \eqref{eq:Sdetection probability}. The procedure to acquire false loads $\boldsymbol{P}_{\text{Atk}}$ is summarized in Alg. \ref{alg:randomAtk}. Varying the attack hour $h$, load shift $\tau$, and number of attacked loads $K$, we can find feasible $\boldsymbol{\Gamma}$ to obtain $\boldsymbol{\gamma}$ using \eqref{eq:randomGamma}, and subsequently create an arbitrary number of false loads $\boldsymbol{P}_{\text{Atk}}$ using \eqref{eq:LRAtk1}. Note that for specific combinations of $h, \tau, K$, and $\mathcal{K}$, sometimes no feasible $\boldsymbol{\Gamma}$ can be found, but we can simply re-run Alg.\ref{alg:randomAtk} with different inputs. Since \eqref{eq:randomGamma} is drawing $\boldsymbol{\gamma}$ randomly from a Gaussian distribution, the resulting real load shift $\tau_r$ of $\boldsymbol{P}_{\text{Atk}}$ may be different than the input $\tau$. We keep drawing $\boldsymbol{\gamma}$ until $\tau_r\le\tau$. The false loads created are then used to generate data samples to train and test the SVM detector. 
	
	\begin{algorithm}[tbh]
		\protect\caption{Generating random LR attack false loads} \label{alg:randomAtk}			
		\textbf{Input:} $h$, $K$, $\tau$ \\
		\textbf{Output:} $\boldsymbol{P}_{\text{Atk}}$
		\begin{enumerate}
			\item Obtain the true loads $\boldsymbol{P}$ at hour $h$.
			\item Randomly select $K$ loads to attack and let $\mathcal{K}$ denote the set of indices of the attacked loads.
			\item Find a $\boldsymbol{\Gamma}$ satisfying \eqref{eq:Gammakk}, \eqref{eq:sumGamma} and \eqref{eq:Sdetection probability} with $\tau, K, \mathcal{K}$, and $\boldsymbol{P}$. This can be done by solving a semidefinite program with arbitrary objective, constrained by \eqref{eq:Gammakk}, \eqref{eq:sumGamma} and \eqref{eq:Sdetection probability}. If no feasible $\boldsymbol{\Gamma}$ can be found, terminate.
			\item Draw the non-zero load changes $\boldsymbol{\gamma}$ from $\mathcal{N}(\boldsymbol{0, \Gamma})$ and obtain false loads $\boldsymbol{P}_{\text{Atk}}$ using \eqref{eq:LRAtk1}.
			\item Calculate the real load shift $\tau_r$ of $\boldsymbol{P}_{\text{Atk}}$ using \eqref{eq:loadshift}. If $\tau_r>\tau$, go to step 4). Otherwise, terminate. 
		\end{enumerate} 
	\end{algorithm} 
	
	\section{Numerical Results \label{sec:Results}}
	We use the publicly available PJM zonal hourly metered load data \cite{PJM2019} from 2015 through 2018 for 20 transmission zones as the historical data to train and test our LR attack detection framework. In order to conveniently create intelligently designed LR attacks as described in Section \ref{sec:designedAtk}, we map each zone to a load bus in the IEEE 30-bus system, leveraging the fact that there are 20 loads in this system. The mapping relationship is adopted from \cite{Pinceti2018}, and all load values are multiplied by a scaling factor of $1.308\times 10^{-3}$ to obtain a system with moderate amount of congestion. Table \ref{tab:mapping} describes the mapping rules between load indices, PJM zones, and bus indices. The SVR and SVM models are implemented in Python using the Scikit-learn package \cite{sklearn}. The random, CM and LO attack creation are implemented in Matlab with solver Gurobi. All experiments are conducted on a 2.7 GHz CPU with 32 GB RAM.
	
	\begin{table}[h]
		\renewcommand{\arraystretch}{1.3}
		\protect\caption{Mapping rules between load indices, PJM zones, and bus indices\label{tab:mapping}}
		\vspace{0.3cm}
		\centering
		\begin{tabular}{!{\vrule width 1.5pt} c|c|c !{\vrule width 1.5pt} c|c|c !{\vrule width 1.5pt}}
			\hline
			Load & Zone & Bus & Load & Zone & Bus \tabularnewline
			\hline
			1 & DOM & 2 & 11 & PL & 17  \tabularnewline
			\hline 
			2 & AE & 3 & 12 & PN & 18  \tabularnewline
			\hline 
			3 & JC & 4 & 13 & PE & 19  \tabularnewline
			\hline 
			4 & CE & 7 & 14 & RECO & 20  \tabularnewline
			\hline 
			5 & AEP & 8 & 15 & ATSI & 21  \tabularnewline
			\hline 
			6 & DPL & 10 & 16 & DUQ & 23  \tabularnewline
			\hline 
			7 & PS & 12 & 17 & BC & 24  \tabularnewline
			\hline 
			8 & DEOK & 14 & 18 & ME & 26  \tabularnewline
			\hline 
			9 & PEP & 15 & 19 & EKPC & 29  \tabularnewline
			\hline 
			10 & DAY & 16 & 20 & AP & 30  \tabularnewline
			\hline 
		\end{tabular}
		\vspace{-0cm}
	\end{table}
	
	\subsection{The SVR Load Predictor Performance \label{sec:SVR_Perform}}
	In this section, we provide details on training and testing the SVR load predictor. As mentioned above, given the hourly load data we have, our SVR load predictor aims to accurately predict the load values at hour $h+1$ when the current hour is $h$. The features we use include time information and historical load values up to hour $h$. We select month ($mo$), hour ($hr$), and weekday/weekend ($wd$) as the time information features, $\boldsymbol{t}=[mo, wd, hr]$. Note that $hr$ here is the wall clock time, for example, $hr=14$ for 2 PM, and is different than $h$, which is a unique point in time. Here we only distinguish between weekdays and weekends since loads tend to be similar during weekdays, \textit{i.e.,} $wd=1$ for weekdays and $wd=2$ for weekends. The temporal correlation of each load is captured by including its historical values, at hour $h$ and $s$ previous hours; and at hour $hr$ and $hr+1$ of $d$ previous days, as features. For load $i$, the load value features $\boldsymbol{f}_i$ are given by
	\begin{flalign}
	\boldsymbol{f}_i = [&{P}_i^{h}, {P}_i^{h-1},...,{P}_i^{h-s},{P}_i^{h-24d}, \notag\\
	&{P}_i^{h-24d+1},...,{P}_i^{h-24},{P}_i^{h-23}]. \label{eq:x_j-feature}
	\end{flalign}
	To capture the spatial correlations, we concatenate the load value features of all the loads. 
	
	The multi-output SVR load predictor is achieved by solving one SVR optimization problem \eqref{eq:SVR} for each load. In our experiments, we trained three SVR models to justify the contribution of capturing spatial correlations, as well as to see the influence of different selected features. Model 1 predicts each load using only time information $\boldsymbol{t}$ and its own load value features. A data sample used in Model 1 to predict load $i$ is given by 
	\begin{flalign}
	\boldsymbol{x}_{j,i}=[\boldsymbol{t}, \boldsymbol{f}_i] \forall i. \label{eq:x_j_1}
	\end{flalign}
	Model 2 and 3 use $\boldsymbol{t}$ and $\boldsymbol{f}_i, \forall i,$ as features to predict all loads. A data sample in these two models is given by
	\begin{flalign}
	\boldsymbol{x}_j=[\boldsymbol{t}, \boldsymbol{f}_1, \boldsymbol{f}_2, ...\boldsymbol{f}_{n_l}], \label{eq:x_j_2}
	\end{flalign}
	where $n_l$ is the number of loads in the system. In Model 2, $s=3$ and $d=2$; and in Model 3, $s=4$ and $d=3$. The ground truth ${y}_{j,i}={P}_{i}^{h+1}$ is the true load value at hour $h+1$ for load $i$. Table \ref{tab:SVR_models} presents some properties of the three tested SVR models. Comparing Models 1 and 2, we can see the influence of considering spatial correlations in addition to temporal correlations, as these two models use the same temporal features, but Model 2 additionally uses the features of all the loads to capture spatial correlations.
	\begin{table}[h]
		\renewcommand{\arraystretch}{1.3}
		\protect\caption{Statistics of SVR models\label{tab:SVR_models}}
		\vspace{0.3cm}
		\centering
		\begin{tabular}{!{\vrule width 1.5pt} c|c|c|c|c|c !{\vrule width 1.5pt}}
			\hline
			Model & $s$ & $d$ & $m$ & $p$ & Training time (h)\tabularnewline
			\hline
			1 & 3 & 2 & 35011 & 11 &  1.927 \tabularnewline
			\hline 
			2 & 3 & 2 & 35011 & 163 & 4.234  \tabularnewline
			\hline 
			3 & 4 & 3 & 34987 & 223 & 33.324  \tabularnewline
			\hline 
		\end{tabular}
		\vspace{-0cm}
	\end{table}
	
	The dimension of the data matrix $\boldsymbol{X}, m\times p,$ and target value matrix $\boldsymbol{Y}, m\times n_l,$ depend on the values of $s$ and $d$. Derivation of $m$ and $p$ are described in the Appendix. For each model, the training data matrix $\boldsymbol{X}_{\text{train}}$ contains all data from 2015 - 2017, and data in 2018 are used as $\boldsymbol{X}_{\text{test}}$. Each column of $\boldsymbol{X}_{\text{train}}$ is scaled to zero mean and unit variance, and each column of $\boldsymbol{X}_{\text{test}}$ is scaled using the mean and variance of the corresponding column in $\boldsymbol{X}_{\text{train}}$. The same split and scaling are performed on $\boldsymbol{Y}$ to obtain $\boldsymbol{Y}_{\text{train}}$ and $\boldsymbol{Y}_{\text{test}}$ as well. The parameters in training the SVR models are chosen as $\varepsilon=10^{-2}$ and $M=100$. The radial basis function (RBF) kernel 
	\begin{equation}\label{eq:RBF}
	Q(\boldsymbol{x}_i, \boldsymbol{x}_j)=-\sigma \|\boldsymbol{x}_i - \boldsymbol{x}_j\|^2
	\end{equation}
	is used with $\sigma=10^{-2}$. Applying the trained SVR predictor on $\boldsymbol{X}_{\text{train}}$ and $\boldsymbol{X}_{\text{test}}$ yields the predicted loads $\hat{\boldsymbol{Y}}_{\text{train}}$ and $\hat{\boldsymbol{Y}}_{\text{test}}$, respectively.
	
	Two metrics are used to evaluate the performance of the SVR load predictor, namely root mean square error (RMSE) and mean absolute percentage error (MAPE). RMSE measures the square root of the average squared error for each load, and hence the unit is MW. MAPE measures on average how much the predicted loads deviate from their true values in percentage. These metrics for each load $i$ are calculated as
	\begin{flalign}
	&\text{RMSE}_{\text{train},i} =\sqrt{\frac{1}{m}\sum_{j=1}^{m}(\boldsymbol{Y}_{\text{train},i,j}-\hat{\boldsymbol{Y}}_{\text{train},i,j})^2}\label{eq:RMSE}\\
	&\text{MAPE}_{\text{train},i} =\frac{1}{m}\sum_{j=1}^{m}\left\vert\frac{\boldsymbol{Y}_{\text{train},i,j}-\hat{\boldsymbol{Y}}_{\text{train},i,j}}{\boldsymbol{Y}_{\text{train},i,j}}\right\vert\label{eq:MAPE}
	\end{flalign}
	where $\boldsymbol{Y}_{\text{train},i}$ is the $i^\text{th}$ column of $\boldsymbol{Y}_{\text{train}}$, and $\bar{\boldsymbol{Y}}_{\text{train},i}$ is its mean. These metrics are similarly applied on $\boldsymbol{Y}_{\text{test}}$ to evaluate the performance of the SVR load predictor on testing data. 
	\begin{figure}[h]
		\centering{}\includegraphics[scale=0.52]{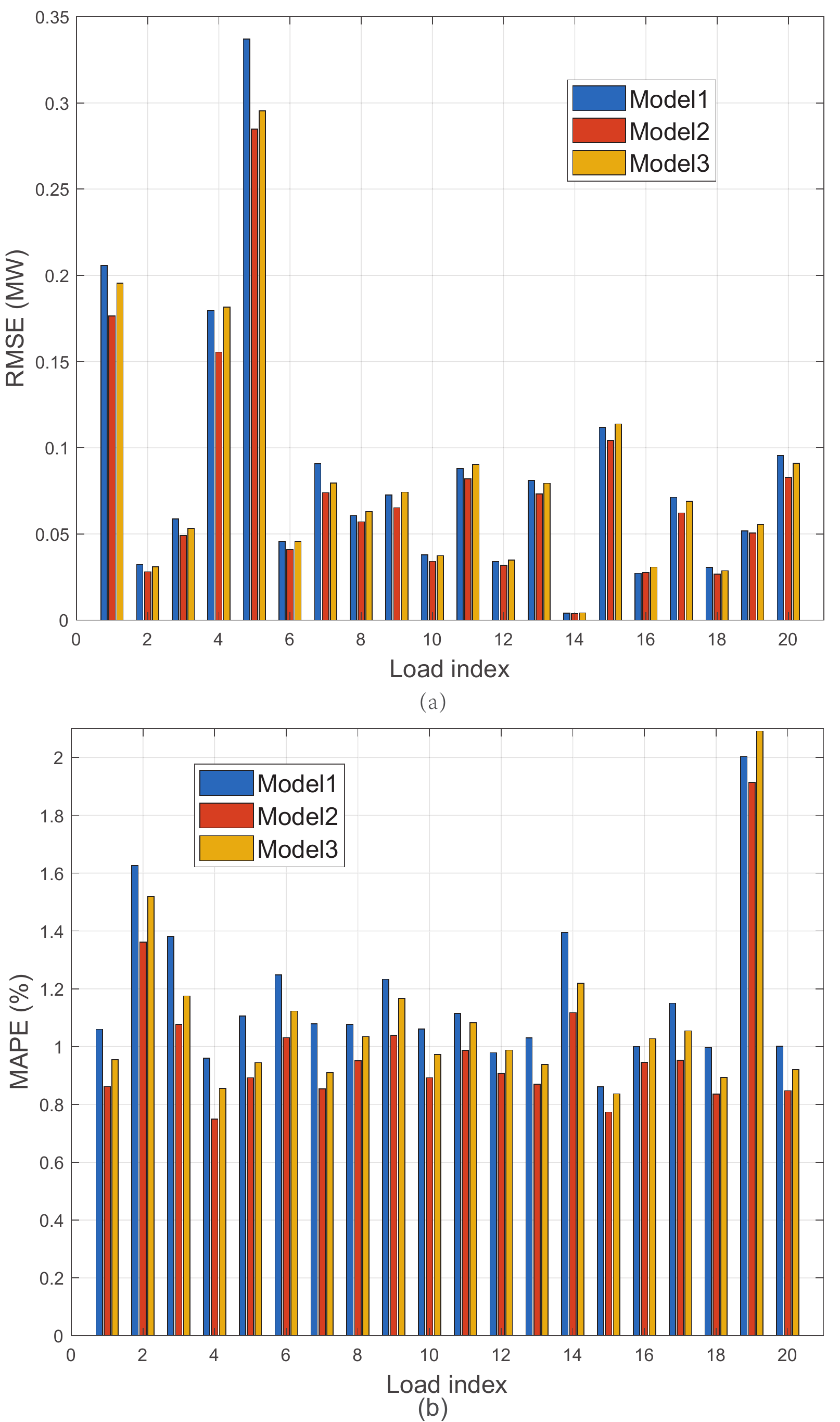}\protect\protect\caption{Performance of the SVR models under two metrics: (a) RMSE, and (b) MAPE. Model 1 does not capture spatial correlations. Model 2 uses temporal features of 3 previous hours and 2 previous days. Model 3 uses temporal features of 4 previous hours and 3 previous days. Both Models 2 and 3 capture spatial correlation. \label{fig:SVR_Performance}}
	\end{figure}
	
	Figures \ref{fig:SVR_Performance} illustrates the RMSE and MAPE for the SVR models. RMSE values largely depend on the load values itself, for example, load 5 has the largest RMSE value because it is the biggest load in the system. From Figure \ref{fig:SVR_Performance}(b) we can see that the MAPE for most loads are around 1\%, and MAPE for load 19, the most difficult load to predict, is around 2\%. Comparing these quantities for Models 1 and 2, we can see that they are both smaller for Model 2. Recall that the difference between Models 1 and 2 is that Model 2 considers all prior loads, while Model 1 only includes the prior data at the load of interest. This result indicates that considering spatial correlations does improve the performance of the SVR load predictor. Comparing Models 2 and 3, it can be concluded that including too much historical data as features decreases the accuracy of the SVR load predictor. Besides, it can be seen from Table \ref{tab:SVR_models} that using too many features makes it extremely slow in training the SVR model. Thus, in the following sections, Model 2 is adopted to generate predicted loads used by the SVM attack detector. 
	
	\subsection{The SVM Attack Detector Performance on Random Attacks\label{sec:SVM_Perform_random}}
	The outputs of the SVR load predictor are used as input features of the SVM attack detector. Depending on the existence of attack, input data samples of the SVM are given by
	\begin{subequations}\label{eq:u_j}
		\begin{flalign}
		\boldsymbol{u}_j &= [mo,wd,hr,\hat{\boldsymbol{P}}, \boldsymbol{P}], \text{if } v_j=-1, \label{eq:u_j_noAtk}\\ 
		\boldsymbol{u}_j &= [mo,wd,hr,\hat{\boldsymbol{P}}, \boldsymbol{P}_{\text{Atk}}], \text{if } v_j=1, \label{eq:u_j_Atk}
		\end{flalign}
	\end{subequations}
	where $v_j=-1$ indicates that there is no attack, and $v_j=1$ otherwise.
	The predicted loads $\hat{\boldsymbol{P}}$ of $m=35011$ hours, along with their ground truth values $\boldsymbol{P}$ and time information, yield $35011$ \textit{normal} data samples for the SVM detector in the form of \eqref{eq:u_j_noAtk}. The length of each data sample $q=3+20\times 2=43$. The \textit{normal} data matrix $\boldsymbol{U}_{\text{normal}}$ is of size $35011\times 43$. We randomly select 80\% of these vectors for training and the remaining 20\% for testing. We create $10^5$ \textit{attacked} data samples in the form of \eqref{eq:u_j_Atk} using Alg. \ref{alg:randomAtk}, resulting in $\boldsymbol{U}_{\text{attack}}$ of size $10^5 \times 43$ with real load shift $\tau_r$ ranging from 1\% to 20\%. From now on, we omit the subscript in $\tau_r$ for easier presentation.
	
	We obtain different SVM models to compare their performances by varying the penalty factor $C$ and $\tau_{\min}$ (the minimal $\tau$ used in the training data). The normal data in the training data matrix $\boldsymbol{U}_{\text{train}}$ are the same for all models, \textit{i.e.,} the same 80\% of $\boldsymbol{U}_{\text{normal}}$. The attacked data in $\boldsymbol{U}_{\text{train}}$ include 80\% of attacked data samples with $\tau\ge \tau_{\min}$. The testing data $\boldsymbol{U}_{\text{test}}$ consists of the remaining 20\% of attacked data that are not used in training with all load shifts, and are the same for all models. For each model, every column of training data matrix $\boldsymbol{U}_{\text{train}}$ is scaled to zero mean and unit variance, and the same scaling is performed to the testing data. The kernel function used in the SVM detector is also the RBF kernel in the form of \eqref{eq:RBF}, but this time $\sigma$ is calculated as $\sigma =1/q$ (this is the ``\textit{scale}'' option in Scikit-learn).
	
\begin{figure}[h]
		\centering{}\includegraphics[scale=0.44]{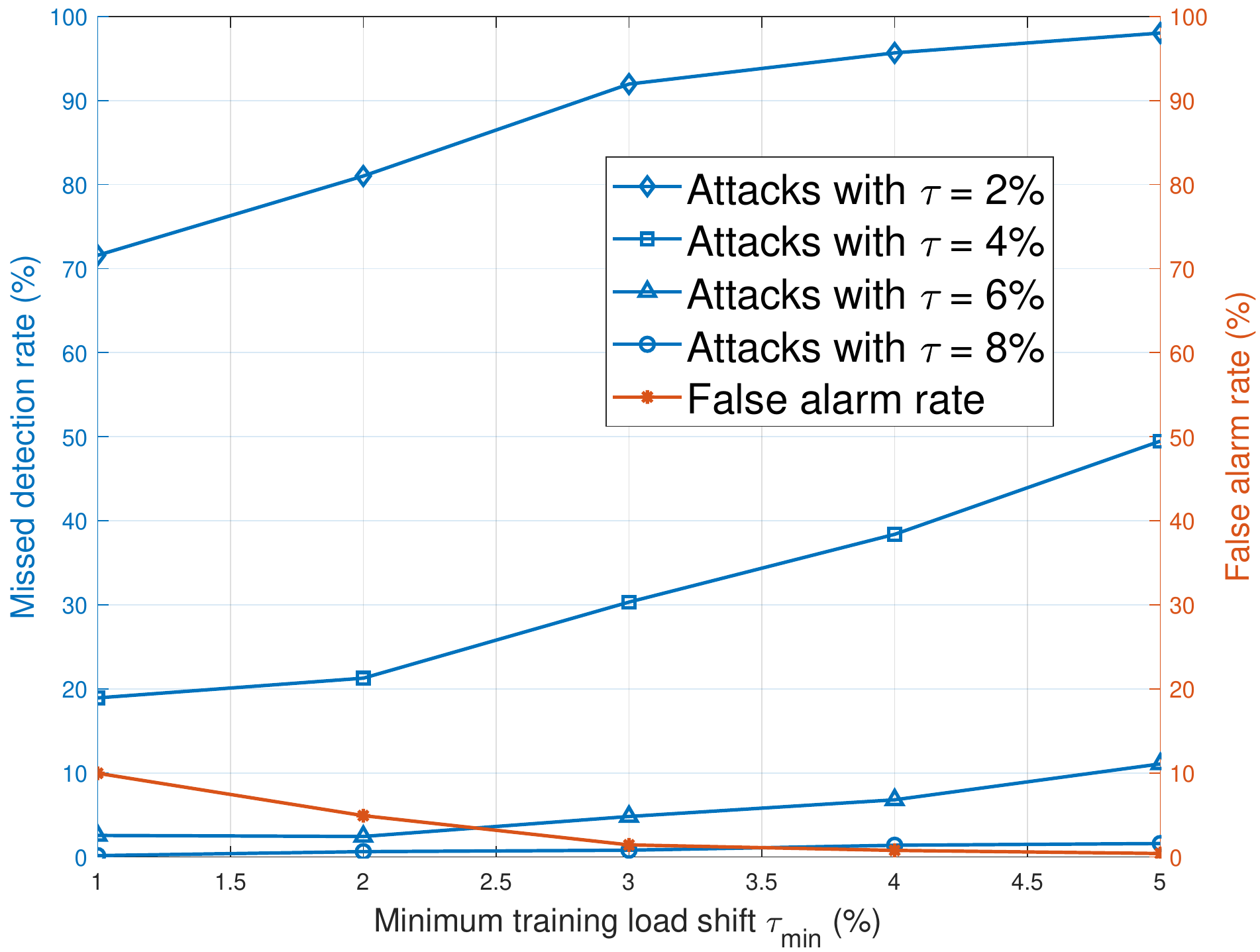}\protect\protect\caption{Effect of minimum training load shift $\tau_{\min}$. False alarm rate and missed detection rate when testing random attacks are each plotted as a function of $\tau_{\min}$. Data is shown for $C=1000$.\label{fig:FAR}}
	\end{figure}

	Figure \ref{fig:FAR} illustrates the effect of $\tau_{\min}$ on missed detection rate and false alarm rate. The false alarm rate is calculated by applying the detector on all $m=35011$ normal data samples, including both training and testing. The parameter $C$ is fixed at $1000$. $\tau_{\min}$ controls the amount of attacked training data. For instance, if $\tau_{\min}=3\%$, $\boldsymbol{U}_{\text{train}}$ contains 80\% of attacks with $\tau\ge 3\%$, but does not contain any attack with $\tau<3\%$. Intuitively, attacks with higher $\tau$ are further away from the normal data than those with lower $\tau$. Thus, a detector trained with a low $\tau_{\min}$ will have a high false alarm rate, as the SVM is trying to find a decision boundary between normal data and attacks with small load shift. However, it should perform better in detecting attacks with small $\tau$ than detectors trained with large $\tau_{\min}$. In Figure \ref{fig:FAR}, the blue lines indicate the missed detection rate of attacks with certain load shift $\tau$, and the red line shows the false alarm rate. It can be seen that as $\tau_{\min}$ increases, the false alarm rate decreases, but the missed detection rate increases for attacks with small load shifts. This observation justifies the intuition discussed above, indicating that $\tau_{\min}$ is indeed a trade-off between false alarm rate and detection probability for small attacks. Note that for attacks with large $\tau$, the effect of $\tau_{\min}$ is insignificant. For testing attacks with extremely small $\tau$, the missed detection rates are very high even with small $\tau_{\min}$, because these attacks are in principle very difficult to detect. However, these attacks are also unlikely to cause severe consequences. From Figure \ref{fig:FAR}, we can see that $\tau_{\min}=3\%$ is a good choice for our dataset. 
	
\begin{figure}[h]
		\centering{}\includegraphics[scale=0.46]{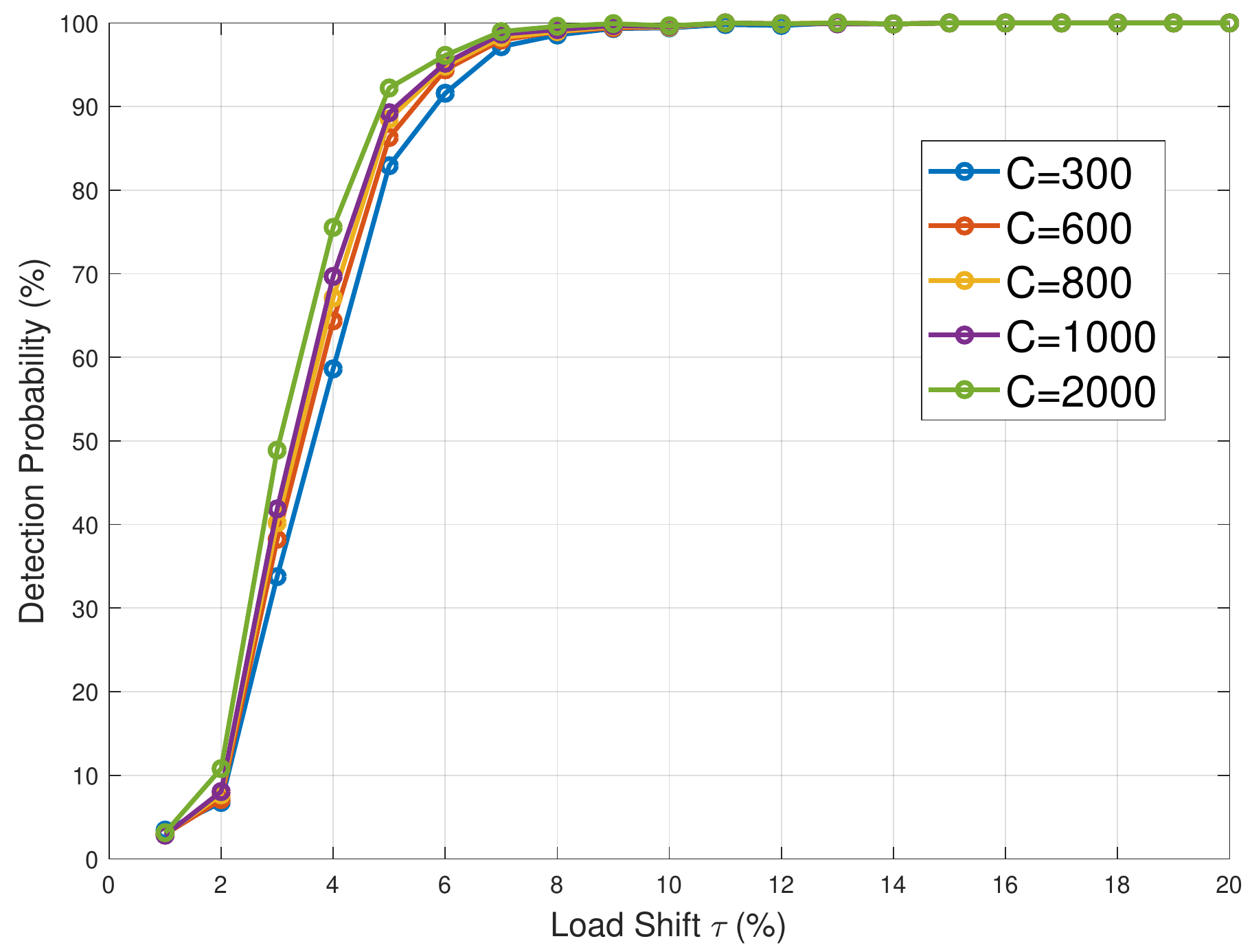}\protect\protect\caption{Effect of outlier penalty factor $C$ on testing random attack detection probability. Data is shown for $\tau_{\text{min}}=3\%$. \label{fig:test3}}
	\end{figure}

	The parameter $C$ trades off misclassification of training examples against simplicity of the decision boundary. A small $C$ makes the decision boundary smooth, while a large $C$ aims at classifying all training samples correctly. Therefore, detector with large $C$ is expected to have a better performance. However, a large $C$ allows for fewer outliers, making it harder to solve the SVM optimization problem \eqref{eq:SVM}, so the training time increases. Figure \ref{fig:test3} shows the performance of models trained with different $C$ on testing random attacks while fixing $\tau_{\min}=3\%$. The larger $C$ is, the higher detection probability we can achieve. This model performs well on attacks with large $\tau$, and the detection probability almost achieves 100\% starting at $\tau=7\%$. System operators can similarly vary $\tau_{\min}$ and $C$ to obtain SVM model with satisfactory performance, in terms of false alarm rate and missed detection rate.
	
	\subsection{The SVM Attack Detector Performance on Intelligently Designed LR Attacks\label{sec:SVM_Perform_intelligent}}
	In this section, we evaluate the performance of the trained SVM detector on cost maximization (CM) and line overflow (LO) attacks. According to the previous section, here we choose SVM parameters $C=2000$ and $\tau_{\min}=3\%$ to balance false alarm rate and missed detection. The procedures to generate these attacks are described as follows. On the IEEE 30-bus system, we first perform base case DCOPF for each hour in year 2015 through 2018 using the true loads. At hour $h$, if there are at least 2 lines whose power flows are greater than 80\% of their ratings, we say those lines are \textit{critical lines}, and $h$ is a \textit{critical hour}. The total number of critical hours is found to be 8038. We focus on critical hours because the false loads are likely to cause congestions at those times, which in turn change the generation dispatch to have consequences. For each critical hour, we solve optimization problem \eqref{eq:CMAtk} 20 times to obtain attack vector $\boldsymbol{c}$ fo CM attacks with $\tau=1\%, 2\%,\dots,20\%$. For each critical line, we solve \eqref{eq:LOObj} 20 times to obtain $\boldsymbol{c}$ for LO attacks, also with $\tau=1\%, 2\%,\dots,20\%$. Every non-zero $\boldsymbol{c}$ is used to construct false load vector $\boldsymbol{P}_{\text{Atk}}$ as in \eqref{eq:falseLoads}. If a $\boldsymbol{P}_{\text{Atk}}$ makes the DCOPF infeasible, it is discarded. The total number of false loads for CM attacks and LO attacks are 113031 and 343135, respectively.  
	
	\begin{figure}[h]
		\centering{}\includegraphics[scale=0.49]{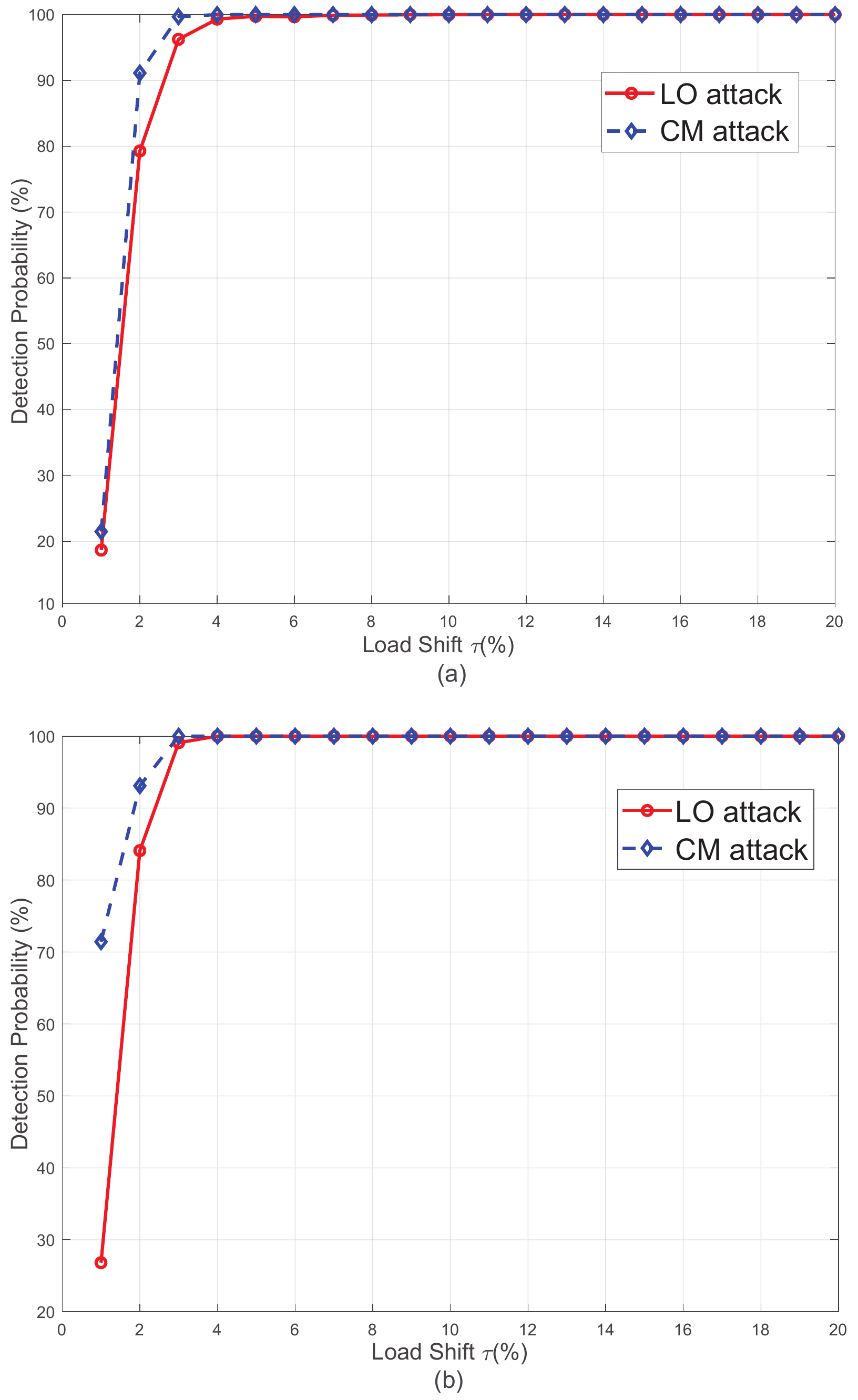}\protect\protect\caption{Detection probability on CM and LO attacks as a function of load shift $\tau$. Subplot (a) is for all attacks, and subplot (b) is only for attacks with consequences. Data is shown for $\tau_{\text{min}}=3\%$ and $C=2000$. \label{fig:Detection_CM_LO}}
	\end{figure}
	Figure \ref{fig:Detection_CM_LO}(a) illustrates the detection probability versus the load shift $\tau$ on CM and LO attacks. For both attacks, the detection probabilities almost achieve 100\% when $\tau\ge 4\%$. For attacks with $\tau=3\%$, the detector performance drops to 97\% for LO attacks, but it is still perfect in detecting CM attacks. Comparing with the performance on random attacks as shown in Figure \ref{fig:test3}, it can be seen that intelligently designed attacks are easier to detect than random attacks. 
	
	Figure \ref{fig:Detection_CM_LO}(b) illustrates the detection probability versus load shift $\tau$ on CM and LO attacks with consequences. CM attacks with consequences are those that increase the operating cost by more than 1\%. LO attacks with consequences are those result in physical overflows. Comparing Figures \ref{fig:Detection_CM_LO}(a) and \ref{fig:Detection_CM_LO}(b), it can be seen that the detector performs even better on attacks with consequences. 
	
	\subsection{Attack Mitigation\label{sec:mitigation}}
	If LR attack is flagged by our detection framework, the simplest way to mitigate the attacks is to temporarily use the loads output by the SVR load predictor for re-dispatching purposes. To test the mitigation performance using this method, we compare the worst consequences of intelligently designed attacks with and without our detection framework. 
	
	In order to obtain the consequences, we run DCOPF three times using different loads. Under normal operation, running DCOPF with true loads $\boldsymbol{P}_{\text{normal}}$ yields the attack-free generation dispatch $\boldsymbol{G}_{\text{normal}}$. Using attacked loads $\boldsymbol{P}_{\text{Atk}}$ to run DCOPF gives attacked dispatch $\boldsymbol{G}_{\text{Atk}}$. Applying $\boldsymbol{G}_{\text{Atk}}$ on true loads $\boldsymbol{P}_{\text{normal}}$ yields attacked line flows $\boldsymbol{P}_{\boldsymbol{L},\text{Atk}}=\boldsymbol{R}(\boldsymbol{G}_{\text{Atk}}-\boldsymbol{P}_{\text{normal}})$. When an attack is detected, the system runs DCOPF using the SVR predicted loads $\boldsymbol{P}_{\text{SVR}}$ and the resulting dispatch is $\boldsymbol{G}_{\text{SVR}}$. The corresponding line flows are given by $\boldsymbol{P}_{\boldsymbol{L},\text{SVR}} = \boldsymbol{R}(\boldsymbol{G}_{\text{SVR}}-\boldsymbol{P}_{\text{normal}})$.
	
\begin{figure}[]
		\centering{}\includegraphics[scale=0.55]{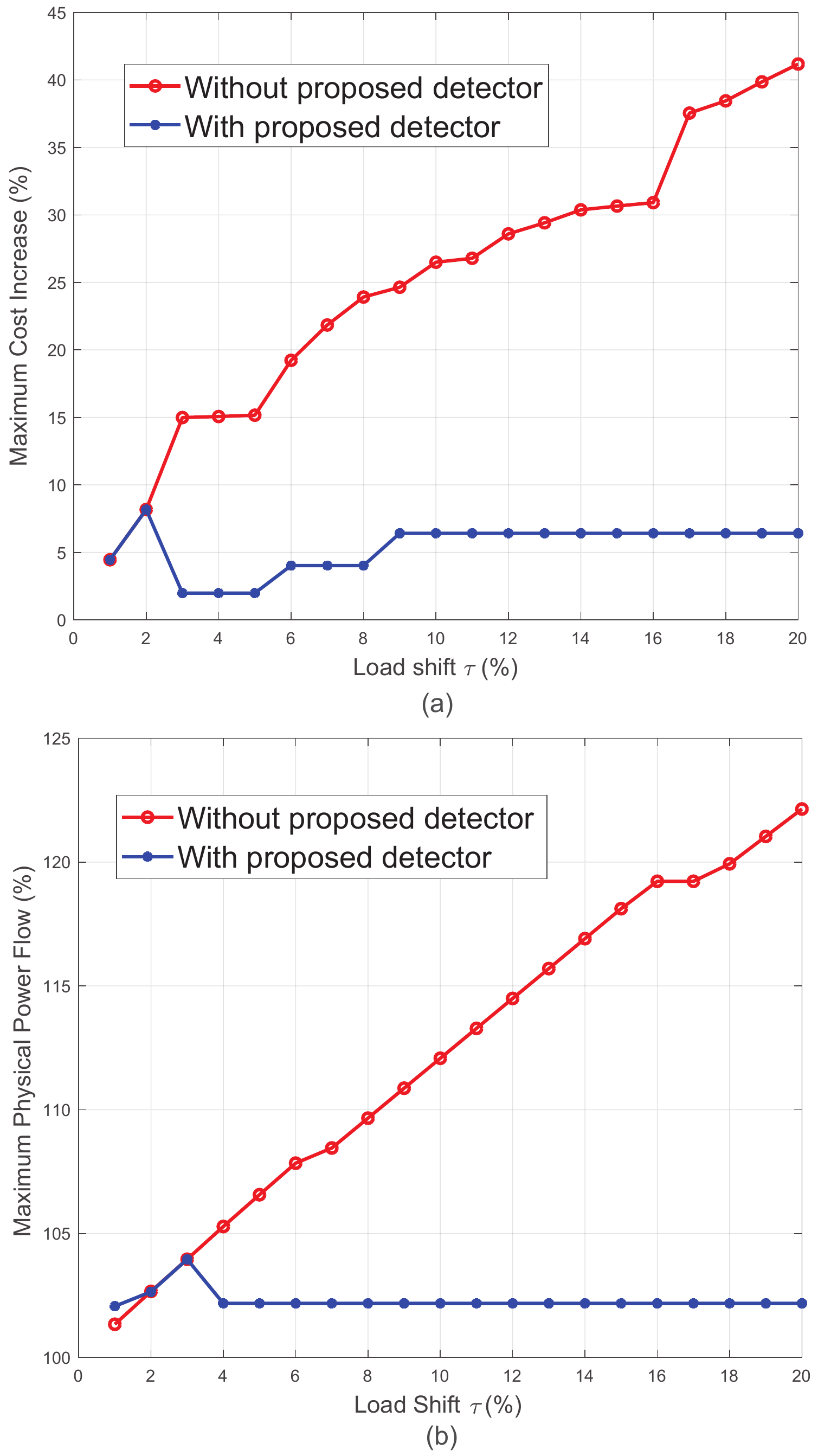}\protect\protect\caption{Mitigation results of (a) CM attacks and (b) LO attacks. For each load shift, the points on the red lines indicate the worst consequence as a result of attack, and the points on the blue lines indicate the worst consequence with our attack detection framework. Points on the blue line are obtained by taking the maximum of two quantities: (i) resulting worst consequence if re-dispatch using SVR predicted loads when attack is flagged; and (ii) the worst attack consequence when the detector fails. \label{fig:Mitigation}}
	\end{figure}

	Figure \ref{fig:Mitigation}(a) illustrates the mitigation results for CM attacks. The word ``maximum'' on the y-axis indicates the worst consequence among all attacks with each load shift $\tau$. The red line indicates the maximum cost increase without using our proposed detection framework, calculated as $\boldsymbol{a}^T(\boldsymbol{G}_{\text{Atk}}-\boldsymbol{G}_{\text{normal}})$ (recall that $\boldsymbol{a}$ is the generation cost vector). When an attack is detected, the resulting cost increase is obtained by $\boldsymbol{a}^T(\boldsymbol{G}_{\text{SVR}}-\boldsymbol{G}_{\text{normal}})$. When the detector fails to detect an attack, the cost increase is the attack consequence $\boldsymbol{a}^T(\boldsymbol{G}_{\text{Atk}}-\boldsymbol{G}_{\text{normal}})$. Thus, for each load shift, if all attacks are detected, the data point on the blue line is given by $\boldsymbol{a}^T(\boldsymbol{G}_{\text{SVR}}-\boldsymbol{G}_{\text{normal}})$. Otherwise, it is $\text{max}[\boldsymbol{a}^T(\boldsymbol{G}_{\text{Atk}}-\boldsymbol{G}_{\text{normal}}), \boldsymbol{a}^T(\boldsymbol{G}_{\text{SVR}}-\boldsymbol{G}_{\text{normal}})]$. Similar procedure is performed to create Figure \ref{fig:Mitigation}(b) for LO attacks. The red line is obtained by taking the maximum $\boldsymbol{P}_{\boldsymbol{L},\text{Atk}}^l$ for each load shift (line $l$ is the target line). The blue line is obtained by $\boldsymbol{P}_{\boldsymbol{L},\text{SVR}}^l$ if all attacks are detected, and $\text{max}[\boldsymbol{P}_{\boldsymbol{L},\text{Atk}}^l, \boldsymbol{P}_{\boldsymbol{L},\text{SVR}}^l]$ otherwise.
	
	%
	
	From Figures \ref{fig:Mitigation}(a), we can see that for load shift $\tau\ge 3\%$, the increases in operation cost are significantly reduced by using SVR predicted loads when an attack is flagged. For LO attacks, the overflows are significantly reduced for load shift $\tau\ge 4\%$. The largest cost increase caused by CM attacks that are not detected is 8.17\% (at $\tau=2\%$), and the largest overflow caused by LO attacks that are not detected is 3.96\% (at $\tau=3\%$). Thus, even though our detector fails to detect a small number of attacks, their consequences are minor. Note that at $\tau=1\%$, using the SVR predicted loads leads to larger overflow due to inaccurate predictions, but the overflow is still very small. Therefore, the consequences of LR attacks can be successfully mitigated using the SVR predicted loads, which gives operators time to take other corrective actions. 
	
	\section{Concluding Remarks \label{sec:conclusion}}
	A machine learning based load redistribution (LR) attack detection framework is proposed. This detection framework consists of a support vector regression (SVR)-based load predictor and a support vector machine (SVM)-based attack detector. The SVR load predictor is trained using features selected from historical load data to capture both spatial and temporal correlations. The prediction results indicate that the SVR load predictor can accurately predict loads at all buses. The SVM attack detector is trained using randomly generated LR attacks, and is shown to be effective in detecting both randomly generated and intelligently designed attacks, especially those with consequences. Using the proposed attack detection framework, system operators can make control decisions based on the predicted loads when attack is flagged to mitigate the consequence of the attacks. It also gives operators time to find the source of the attacks. Future work will include finding attack localization techniques that help system operators identify the loads and/or meters that are modified by the attacker. 
	
	\section*{Acknowledgment}
	This material is based on work supported by the National Science Foundation (NSF) under grant number CNS-1449080, and two grants from the Power System Engineering Research Center (PSERC) S-72 and S-74.
	
	\bibliographystyle{iet}
	\bibliography{Journal_AtkDetection}
	
	\section*{Appendix}
	
	The parameters $s$ and $d$ in \eqref{eq:x_j-feature} determines the dimension of SVR input data matrix $\boldsymbol{X}, m\times p$. For example, for Model 2, $s=3$ and $d=2$, the length of $\boldsymbol{f}_i$ is given by
	\begin{equation}
	n_f = s+1+2d = 8.
	\end{equation} 
	The resulting data sample length $p=3+20\times n_f=163$. Since we use load values of previous $d=2$ days as features, the start hour of our data is 01/03/2015, 0 AM. The end hour is 12/31/2018, 10 PM because for 12/31/2018, 11 PM, we do not have ground truth values of its next hour. In each of the four years, the hour when daylight saving time ends has two load values with identical time stamps, and we approximate the load value at this hour by taking the average of those two values. As a result, the number of data samples for the SVR load predictor is 
	\begin{equation}
	m=(365\times 3+366-d)*24-1-4=35011.
	\end{equation}
	The target values for hour $h$ are the metered loads of the 20 zones at hour $h+1$. Thus, for each data sample of length $p=163$, the SVR outputs a vector of length 20 as prediction. We use the first $26253$ data samples in year 2015 through 2017 to train the SVR load predictor and use the remaining $8758$ data samples in 2018 to test its performance. The resulting training data matrix $\boldsymbol{X}_{\text{train}}$ is of size $26253\times 163$, training target value matrix $\boldsymbol{Y}_{\text{train}}$ is of size $26253\times 20$, testing data matrix $\boldsymbol{X}_{\text{test}}$ is of size $8758\times 163$, and the testing target value matrix $\boldsymbol{Y}_{\text{test}}$ is of size $8758\times 20$. The dimensions of these matrices for other models can be similarly determined.
	
\end{document}